# A Potential Correlation Between the Temperature of the Pacific Ocean and Data from Google Trends May Yield a Warning Sign for the Outbreak of Zika

Raúl Isea

Institute of Advanced Studies – IDEA, Hoyo de la Puerta, Baruta, Venezuela

### Email address

risea@idea.gob.ve

### Abstract

There has been a large number of reported cases of the occurrence of Zika in different countries in 2016 and it is necessary to develop an early warning system to initiate preventive campaigns against the disease. A potential early warning system based on the rise in ocean temperature of the Pacific Niño Index is proposed. The efficacy is verified using data for the outbreak in Colombia as obtained from Google Trends.

## 1. Introduction

Zika has the gene *Flavivirus*, the same gene that is contained in Dengue and Chikungunya. It was first isolated in the Zika Forest of Uganda in 1947 and it has often been confused with dengue. It was in 2007 that the first case of Zika was reported in Africa and Asia in the Federated States of Micronesia [1]. Recently, Zika has been reported as an emerging pathogen and a new threat for Latin America [2]. At the present time, there is no effective treatment against the disease.

For that reason, it is important to develop an early warning system for detecting and preventing the disease epidemics throughout the world. The epidemiological alert system usually suffers a slight delay from the initial detection of the outbreak. For example, Argentina has had the last severe dengue outbreaks in 2009 and later studies indicated that it was linked to factors such as climate change, urbanization and the decreased use of pesticides that reliably kill the mosquito vector [3].

Following these results, we propose a possible correlation between the temperature of the Pacific Ocean and an outbreak of Zika. Preliminary results will be compared with data resulting from an outbreak of Zika in Colombia. This idea was originated by Isea *et al* [4]. Other authors published the possible relation between the vector and the climate, see for example Campbell [5], Rosà *et al* [6], Gong *et al* [7], Hafez [8] and the references cited in these papers.

On the other hand, it is also worth mentioning the available tools for the online detection and quantification of epidemics [9]. By means of the analysis of health care data and statistical periodic regression models, it is possible to detect outbreaks and estimate epidemic burdens if a theoretical threshold is reached in the periodic baseline, which requires of a period of several years of calculated.

Using all these ideas, this paper suggests the possibility to develop an early warning system against the disease Zika by monitoring the difference of the temperatures of the Oceanic Niño Index (ONI) and the data obtained using Google Trends. The efficacy is tested using data for the outbreak in Colombia.

## 2. Model

The variable climate change to be used in this work is obtained from the Oceanic Niño Index (abbreviated as ONI and available at http://www.cpc.ncep.noaa.gov) which represents the deviation from the average surface temperature of the Pacific Ocean. The surface temperature of the ocean that is recorded in the equatorial Pacific Ocean (located at 5N-5S, 120-170W), known as El Niño 3.4. The Oceanic Niño 3.4 index, is one of the largest databases that measures the climate phenomenon. The next step was to obtain a record of cases of Zika during October 2015 until March 2016 which was obtained from the Colombia National Institute of Health epidemiological bulletin (available at www.paho.org/). Finally, we employed the data obtained using Google Trends (available at https://www.google.com/trends), and determined the possible correlation among them employing the Spearman correlation coefficient [10].

## 3. Results

The results are displayed in two figures. Figure 1 displays the difference temperature of the ONI from October 2014 until March 2016 in red and the data obtained from Google Trends concerning the disease Zika in Colombia during the same period of time in blue. From this data, we determined the Spearman correlation coefficient (R) that is based on the null hypothesis between the ONI data and the Google Trends results. It is interesting to note that the resulting value is R = 0.71. This indicates that there is a modest positive correlation. This implies that there is a tendency for the Google Trends and the ONI data to be somewhat correlated (a value of R = 1 indicates 100% correlation).

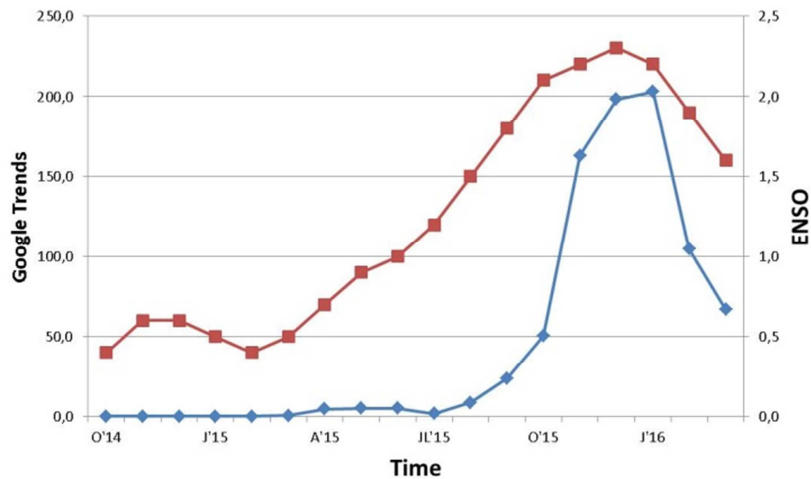

*Figure 1.* *The index of Google Trends in Colombia (blue color) from October 2014 until March 2016, and the temperature of the Pacific Ocean using ENSO in red color.*

Figure 2 shows the number of suspected cases of Zika in Colombia during the time interval September 2015 to March 2016 in blue. Data obtained from ONI is shown in red in Figure 1. This graph indicates that there is a possible correlation between these variables with R = 0.15. It is interesting to note that when the numerical values of the ONI decreases, the cases of Zika in Colombia also decrease.

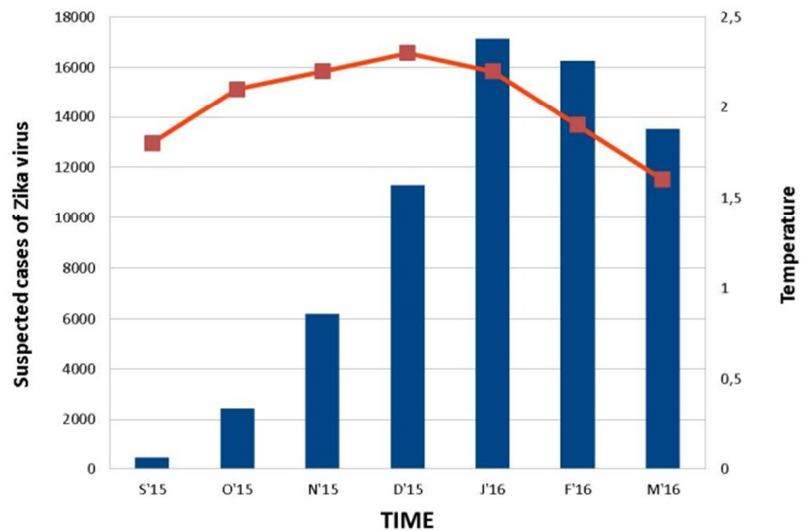

*Figure 2.* *Number of suspected cases of Zika in Colombia from September 2015 to March 2016 in bars, and the red line the Oceanic Niño Index temperature (see text for details).*

## 4. Conclusion

From the results obtained in this paper, we suggest that it is reasonable to include the Oceanic Niño Index (ONI) as an early indicator of a possible Zika outbreak in Colombia. In fact, an increase in the ocean temperature could be associated with an increase in the number of cases of Zika. For this reason, it is necessary to develop tools for the online early detection and quantification of possible Zika epidemics.

## Acknowledgements

The author wishes to express his sincere thanks to Prof. Karl E. Lonngren for his unconditional help and the comments concerning the manuscript.

## References

[1] Dick GW, Kitchen SF, Haddow AJ (1952) Zika virus. I. Isolations and serologicalspecificity. *Trans R Soc Trop Med Hyg* 1952; 46: 509-520.

[2] Alfonso J Rodríguez-Morales. Zika: the new arbovirus threat for Latin America. *J Infect Dev Ctries* 2015; 9(6): 684-685.

[3] Gupta, Ritu; Tiwari, Rishi; Mueen, Ammed K. K. Argentina's dengue-fever outbreak reaches capital. Nature 2009; 458 (7240), 819.

[4] Isea R, Mayo-García R, and Lonngren, K. E. A potential correlation between dengue and the temperature of the Pacific Ocean yielding an additional warning sign. *Environmental Science: An Indian Journal* 2012, 7 (11): 395-396.

[5] Campbell LP, Luther C, Moo-Llanes D, Ramsey JM, Danis-Lozano R, Peterson AT. Climate change influences on global distributions of dengue and chikungunya virus vectors. *Phil. Trans. R. Soc*. B 2015; 370: 20140135.

[6] Rosà R, Marini G, Bolzoni L, Neteler M, Metz M, Delucchi L, et al. Early warning of West Nile virus mosquito vector: climate and land use models successfully explain phenology and abundance of Culex pipiens mosquitoes in north-western Italy. *Parasit Vectors*. 2014; 12 (7): 269.

[7] Gong H, DeGaetano AT, Harrington LC. Climate-based models for West Nile Culex mosquito vectors in the Northeastern US. *Int J Biometeorol*. 2011; 55 (3): 435–446.

[8] Hafez, Y. Study on the Relationship between the Oceanic Nino Index and Surface Air Temperature and Precipitation Rate over the Kingdom of Saudi Arabia. Journal of Geoscience and Environment Protection. 2016; 4: 146-162.

[9] Pelat C, Boëlle PY, Cowling BJ, Carrat F, Flahault A, Ansart S, Valleron AJ. Online detection and quantification of epidemics. *BMC Med Inform Decis Mak*. 2007; 15 (7): 29.

[10] Myers JL, Well AD. Research Design and Statistical Analysis (2nd ed.). Lawrence Erlbaum. 2003; p. 508. ISBN 0-8058-4037-0.